\documentclass[12pt]{article}
\usepackage{graphicx}
\usepackage[cp1251]{inputenc}

\begin{document}
 \noindent {\footnotesize\it Astronomy Letters, 2019, Vol. 45, No 3, pp. 151--162.}
 \newcommand{\dif}{\textrm{d}}

 \noindent
 \begin{tabular}{llllllllllllllllllllllllllllllllllllllllllllll}
 & & & & & & & & & & & & & & & & & & & & & & & & & & & & & & & & & & & & & \\\hline\hline
 \end{tabular}

  \vskip 0.5cm
  \centerline{\bf\large Kinematics of the Galaxy from a Sample of Young Open Star}
  \centerline{\bf\large Clusters with Data from the Gaia DR2 Catalogue}
   \bigskip
  \bigskip
  \centerline
 {V.V. Bobylev\footnote [1]{e-mail: vbobylev@gaoran.ru} and A.T. Bajkova}
  \bigskip

  \centerline{\small\it Pulkovo Astronomical Observatory, Russian Academy of Sciences,}

  \centerline{\small\it Pulkovskoe sh. 65, St. Petersburg, 196140 Russia}
 \bigskip
 \bigskip
 \bigskip

 {
{\bf Abstract}---We have selected a sample of 326 young $(\log
t<8)$ open star clusters with the proper motions and distances
calculated by various authors from Gaia DR2 data. The mean values
of their line-of-sight velocities have also been taken from
various publications. As a result of our kinematic analysis, we
have found the following parameters of the angular velocity of
Galactic rotation:
 $\Omega_0=29.34\pm0.31$ km s$^{-1}$ kpc$^{-1}$,
 $\Omega'_0=-4.012\pm0.074$ km s$^{-1}$ kpc$^{-2}$, and
 $\Omega''_0=0.779\pm0.062$ km s$^{-1}$ kpc$^{-3}$. The circular rotation velocity of
the solar neighborhood around the Galactic center is
$V_0=235\pm5$~km s$^{-1}$ for the adopted Galactocentric distance
of the Sun $R_0=8.0\pm0.15$~kpc. The amplitudes of the tangential
and radial velocity perturbations produced by the spiral density
wave are
 $f_\theta=3.8\pm1.2$~km s$^{-1}$ and
 $f_R=4.7\pm1.0$~km s$^{-1}$, respectively; the perturbation wavelengths are
 $\lambda_\theta=2.3\pm0.5$~kpc and
 $\lambda_R=2.2\pm0.5$~kpc for the adopted four-armed
spiral pattern. The Sun's phase in the spiral density wave is
close to $\chi_\odot=-120\pm10^\circ.$
  }


 \subsection*{INTRODUCTION}
Open star clusters (OSCs) play an important role for studying the
Galaxy and its subsystems, because the mean values of a number of
kinematic and photometric parameters derived from them are highly
accurate. OSCs are used as a tool for studying the properties of
the Galactic thin and thick disks, their dynamical and chemical
evolution, the spiral structure, the star formation processes,
establishing the distance scale, etc.

The second Gaia data release (Gaia DR2) was published in April
2018 (Brown et al. 2018; Lindegren et al. 2018), while the third
data release is scheduled to be issued in mid-2020. The Gaia DR2
catalogue contains the trigonometric parallaxes and proper motions
of $\sim$1.7 billion stars. The derivation of their values is
based on the orbital observations performed for 22 months. The
mean errors of the trigonometric parallax and both proper motion
components in this catalogue depend on magnitude. For example, the
parallax errors lie in the range 0.02--0.04 mas for bright stars
$(G<15^m)$ and are 0.7 mas for faint stars $(G=20^m).$ For quite a
few (more than 7 million) stars of spectral types F--G--K their
line-of-sight velocities were determined with a mean error of
$\sim$1 km s$^{-1}$.

Using highly accurate Gaia DR2 data has allowed one to derive new
mean values of the kinematic parameters for quite a few OSCs
(Babusiaux et al. 2018; Kuhn et al. 2019; Cantat-Gaudin et al.
2018), to study the spatial and intrinsic kinematic properties of
a number of young stellar associations (Zari et al. 2018;
Franciosini et al. 2018; Roccatagliata et al. 2018; Kounkel et al.
2018) and OSCs (Soubiran et al. 2018; Dias et al. 2018) close to
the Sun with unprecedented detail, to detect new OSCs (Beccari et
al. 2018), and to study the fine structure of the
Hertzsprung--Russell diagram (Babusiaux et al. 2018) important for
refining the empirical isochrones and the evolutionary processes,
which must result in a deeper understanding of the physics of
stars.

At relative parallax errors for stars from the Gaia DR2 catalogue
less than 10\% the radius of the solar neighborhood with these
stars is $\sim$3 kpc (Fig. 1 in Xu et al. (2018)). This allows one
to cover almost the entire Local Arm and to reach the edges of the
Perseus and Carina--Sagittarius arms and to determine the
parameters of the spiral structure.

Previously (Bobylev and Bajkova 2018), based on a sample of
$\sim$500 OB stars with proper motions and parallaxes from the
Gaia DR2 catalogue, we refined the Galactic rotation parameters
and the parameters of the spiral density wave. One might expect
that, given the necessary statistics, a kinematic analysis of OSCs
using the parameters calculated from GaiaDR2 data will allow these
results to be confirmed or even improved, because the velocities
of OSCs are determined with a higher accuracy than are the
velocities of single stars.

The goal of this paper is to refine the rotation parameters of the
Galaxy and its spiral structure using the latest data on OSCs. For
this purpose, we use the mean proper motions and parallaxes of
OSCs calculated by various authors exclusively from Gaia DR2 data,
while the mean line-of-sight velocities of these OSCs were derived
mostly from ground-based observations, although there are cases
where they were determined from Gaia DR2 data.

 \section*{DATA}
 \subsection*{Proper Motions and Line-of-Sight Velocities of OSCs}
The main source of the mean proper motions and parallaxes
calculated from Gaia DR2 data for us was the paper by
Cantat-Gaudin et al. (2018), where these quantities were
determined for 1229 OSCs. The parameters of several more OSCs were
taken from Babusiaux et al. (2018), where they were calculated
exclusively from Gaia DR2 data based on a large number of most
probable cluster members.

We took the mean heliocentric line-of-sight velocities of OSCs
mostly from the MWSC (Milky Way Star Clusters) catalogue
(Kharchenko et al. 2013) and, in several cases, from Kuhn et al.
(2018), Babusiaux et al. (2018), Casamiquela et al. (2016), Conrad
et al. (2014), and Mermilliod et al. (2008). Soubiran et al.
(2018) showed that there is good agreement between the
line-of-sight velocities of OSCs calculated only from Gaia DR2
data and those from the MWSC catalogue.

\begin{figure}[t]
{\begin{center}
   \includegraphics[width=0.5\textwidth]{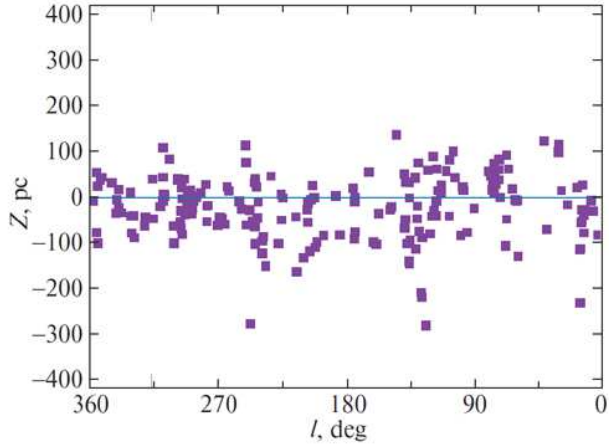}
   \caption{
 Positions of young $(\log t<8)$ OSCs relative to the Galactic plane.
  } \label{f-lz}
\end{center}}
\end{figure}
\begin{figure}[t]
{\begin{center}
   \includegraphics[width=0.99\textwidth]{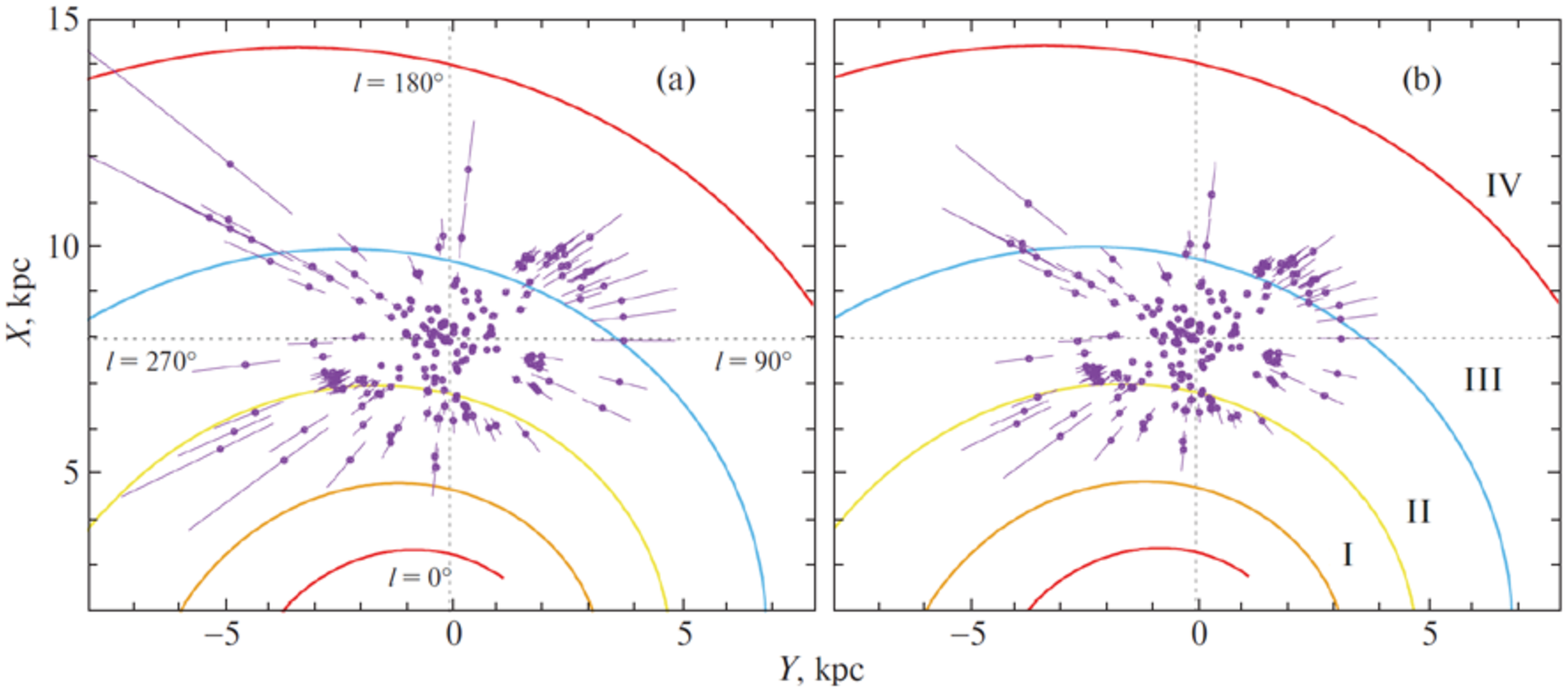}
 \caption{
Distribution of young ($\log t<8$) OSCs whose distances were
calculated using the original parallaxes from the Gaia DR2
catalogue (a) and with the correction $\Delta\pi=0.050$ mas (b) on
the Galactic $XY$ plane; The Sun has coordinates $(X,Y)=(8,0)$
kpc, the four-armed spiral pattern with a pitch angle of
$-13^\circ$ is shown (Bobylev and Bajkova 2014), the spiral arm
segments are numbered by Roman numerals.
  } \label{f-XY}
\end{center}}
\end{figure}

In this paper we consider OSCs with relative parallax errors
$\sigma_\pi/\pi<30\%,$ where the dispersion $\sigma_\pi$ was taken
from column 109 in the catalogue by Cantat-Gaudin et al. (2018).
There are 925 such OSCs of various ages for each of which there
are proper motions and parallaxes. Out of them, 459 OSCs also have
line-of-sight velocity estimates; for these clusters we can
calculate their total space velocities. The last sample contains
211 relatively young OSCs for which $\log t<8.$ Precisely these
OSCs are of greatest interest for studying the Galactic
kinematics, because they belong to the rotating thin disk, are
affected by the spiral density wave, and must have a low residual
velocity dispersion. In this sample the relative parallax errors
for all OSCs do not exceed 30\%. Their distribution on the $l--Z$
plane ($l$ is the Galactic longitude, $Z$ is the coordinate in a
rectangular coordinate system toward the Galactic Pole) is shown
in Fig. 1. As can be seen from the figure, all these OSCs are no
more than 300 pc away from the Galactic plane, i.e., they all
belong to the thin disk. An asymmetry in the distribution of OSCs
relative to the horizontal axis is also clearly seen. This
reflects the well-known fact of the Sun’s elevation above the
Galactic plane. From the data on 211 OSCs we found $Z_\odot=-20\pm
5$~pc. This value is in good agreement with the results of our
analysis of samples of other young thin-disk objects (Bobylev and
Bajkova 2016).

 \subsection*{Correction to the Gaia DR2 Parallaxes}
The presence of a possible systematic offset $\Delta\pi=-0.029$
mas in the Gaia DR2 parallaxes with respect to an inertial
reference frame was first pointed out by Lindegren et al. (2018).
Here the minus means that this correction should be added to the
Gaia DR2 stellar parallaxes to reduce them to the standard. At
present, there are several reliable distance scales a comparison
with which, in the opinion of their authors, allows the
systematics of the Gaia trigonometric parallaxes to be controlled.
Arenou et al. (2018) compared the Gaia DR2 parallaxes with 29
independent catalogues that confirm the presence of an offset in
the Gaia DR2 parallaxes $\Delta\pi\sim-0.03$ mas.

Stassun and Torres (2018) found the correction
$\Delta\pi=-0.082\pm0.033$~mas by comparing the parallaxes of 89
detached eclipsing binaries with their trigonometric parallaxes
from the Gaia DR2 catalogue. These stars were selected from
published data using very rigorous criteria imposed on the
photometric parameters. As a result, the relative errors in the
stellar radii, effective temperatures, and bolometric
luminosities, from which the distances are estimated, do not
exceed 3\%.

Bobylev (2019) obtained an estimate of
$\Delta\pi=-0.038\pm0.046$~mas from a comparison of 88 radio stars
whose trigonometric parallaxes were measured by various authors by
means of VLBI with the Gaia DR2 catalogue. It is well known that
this method allows the stellar parallaxes to be determined with an
error of $\sim10$~$\mu$as. However, so far there are few such
stars and, therefore, the error in the estimate is great.

By comparing the astrometric (Gaia DR2) and photometric parallaxes
of 94 OSCs, Yalyalieva et al. (2018) found the correction
$\Delta\pi=-0.045\pm0.009$~mas. The high accuracy of this estimate
is related to the high accuracy of photometric distance estimates
for OSCs. The data from up-to-date first-class infrared
photometric surveys, such as IPHAS, 2MASS, WISE, and Pan-STARRS,
were invoked for this purpose.

Riess et al. (2018) obtained an estimate of
$\Delta\pi=-0.046\pm0.013$~mas based on a sample of 50 long-period
Cepheids when comparing their parallaxes with those from the Gaia
DR2 catalogue. The photometric parameters of these Cepheids
measured from the Hubble Space Telescope were used.

By comparing the distances of $\sim$3000 stars from the APOKAS-2
catalogue (Pinsonneault et al. 2018) belonging to the red giant
branch with the Gaia DR2 data, Zinn et al. (2018) found the
correction $\Delta\pi=-0.053\pm0.003$~mas. These authors also
obtained a close value by analyzing stars belonging to the so
called red clump, $\Delta\pi=-0.050\pm0.004$~mas. The distances to
such stars were estimated from asteroseismic data. According to
these authors, the parallax errors here are approximately equal to
the errors in estimating the stellar radius and are, on average,
1.5\%. Such small errors in combination with the enormous number
of stars allowed $\Delta\pi$ to be determined with a high
accuracy.

Note also the experiment to compare the distances to OSCs from
various catalogues described in Cantat-Gaudin et al. (2018). It
showed that the correction $\Delta\pi$ differs from that
recommended by Lindegren et al. (2018) and should be close to
$-0.050$~mas.

The listed results lead to the conclusion that the trigonometric
parallaxes of stars from the Gaia DR2 catalogue should be
corrected by applying a small correction. We will be oriented to
the results of Yalyalieva et al. (2018), Riess et al. (2018), and
Zinn et al. (2018), which look most reliable.

Note that two types of distances are given in the catalogue by
Cantat-Gaudin et al. (2018). First, it gives the mean parallaxes
of OSCs calculated from the original trigonometric parallaxes of
probable cluster members that were taken by these authors from the
Gaia DR2 catalogue. In this paper we use precisely these values
(and similar values taken from other authors) to calculate the
distances to OSCs. Second, it gives the distances to OSCs
calculated from the mean parallaxes by adding the correction
$\Delta\pi=0.029$~mas, but these distances are not used here.

Figure 2 shows the distribution of young OSCs whose distances were
calculated both using the original mean parallaxes from the Gaia
DR2 catalogue and by adding the correction $\Delta\pi=0.050$~mas
to these values on the Galactic $XY$ plane. The Roman numerals in
the figure number the following spiral arm segments: Scutum (I) ,
Carina--Sagittarius (II), Perseus (III), and the Outer Arm (IV).
It follows from the figure that the correction affects
significantly the distance calculations for OSCs, especially those
far from the Sun. Note also that the distribution of points in
Fig. 2b agrees better with the above spiral pattern.

 \section*{METHOD}
We know three stellar velocity components from observations: the
line-of-sight velocity $Vr$ and the two tangential velocity
components $V_l=4.74r\mu_l\cos b$ and $V_b=4.74r\mu_b$ along the
Galactic longitude $l$ and latitude $b,$ respectively, expressed
in km s$^{-1}.$ Here, the coefficient 4.74 is the ratio of the
number of kilometers in an astronomical unit to the number of
seconds in a tropical year, and $r=1/\pi$ is the stellar
heliocentric distance in kpc. The proper motion components
$\mu_l\cos b$ and $\mu_b$ are expressed in mas yr$^{-1}.$ The
velocities $U,V,W$ directed along the rectangular Galactic
coordinate axes are calculated via the components $V_r, V_l, V_b:$
 \begin{equation}
 \begin{array}{lll}
 U=V_r\cos l\cos b-V_l\sin l-V_b\cos l\sin b,\\
 V=V_r\sin l\cos b+V_l\cos l-V_b\sin l\sin b,\\
 W=V_r\sin b                +V_b\cos b,
 \label{UVW}
 \end{array}
 \end{equation}
where the velocity $U$ is directed from the Sun toward the
Galactic center, $V$ is in the direction of Galactic rotation, and
$W$ is directed to the north Galactic pole. We can find two
velocities, $V_R$ directed radially away from the Galactic center
and the velocity $V_{circ}$ orthogonal to it pointing in the
direction of Galactic rotation, based on the following relations:
 \begin{equation}
 \begin{array}{lll}
  V_{circ}= U\sin \theta+(V_0+V)\cos \theta, \\
       V_R=-U\cos \theta+(V_0+V)\sin \theta,
 \label{VRVT}
 \end{array}
 \end{equation}
where the position angle $\theta$ obeys the relation
$\tan\theta=y/(R_0-x)$, and $x,y,z$ are the rectangular
heliocentric coordinates of the star (the velocities $U,V,W$ are
directed along the corresponding $x,y,z$ axes), $V_0$ is the
linear rotation velocity of the Galaxy at the solar distance
$R_0.$ The velocities $V_R$ and $W$ are virtually independent of
the pattern of the Galactic rotation curve. However, to analyze
the periodicities in the tangential velocities, it is necessary to
determine a smoothed Galactic rotation curve and to form the
residual velocities $\Delta V_{circ}$.

To determine the parameters of the Galactic rotation curve, we use
the equations derived from Bottlinger’s formulas, in which the
angular velocity $\Omega$ is expanded into a series to terms of
the second order of smallness in $r/R_0:$
\begin{equation}
 \begin{array}{lll}
 V_r=-U_\odot\cos b\cos l-V_\odot\cos b\sin l-W_\odot\sin b\\
    +R_0(R-R_0)\sin l\cos b\Omega^\prime_0
 +0.5R_0(R-R_0)^2\sin l\cos b\Omega^{\prime\prime}_0,
 \label{EQ-1}
 \end{array}
 \end{equation}
 \begin{equation}
 \begin{array}{lll}
 V_l= U_\odot\sin l-V_\odot\cos l-r\Omega_0\cos b\\
    +(R-R_0)(R_0\cos l-r\cos b)\Omega^\prime_0
 +0.5(R-R_0)^2(R_0\cos l-r\cos b)\Omega^{\prime\prime}_0,
 \label{EQ-2}
 \end{array}
 \end{equation}
 \begin{equation}
 \begin{array}{lll}
 V_b=U_\odot\cos l\sin b+V_\odot\sin l \sin b-W_\odot\cos b\\
     -R_0(R-R_0)\sin l\sin b\Omega^\prime_0
  -0.5R_0(R-R_0)^2\sin l\sin b\Omega^{\prime\prime}_0,
 \label{EQ-3}
 \end{array}
 \end{equation}
where $R$ is the distance from the star to the Galactic rotation
axis:
 \begin{equation}
 R^2=r^2\cos^2 b-2R_0 r\cos b\cos l+R^2_0.
 \end{equation}
The quantity $\Omega_0$ is the angular velocity of Galactic
rotation at the solar distance $R_0,$ the parameters
$\Omega^\prime_0$ and $\Omega^{\prime\prime}_0$ are the
corresponding derivatives of the angular velocity, and
$V_0=|R_0\Omega_0|$. As experience shows, to construct a smooth
Galactic rotation curve in the range of distances $R$ from 2 to 12
kpc, it will suffice to know two derivatives of the angular
velocity, $\Omega^\prime_0$ and $\Omega^{\prime\prime}_0$. Note
that the velocities $V_R$ and $\Delta V_{circ}$ must be freed from
the peculiar solar velocity $U_\odot,V_\odot,W_\odot$.

A number of studies devoted to determining the mean distance from
the Sun to the Galactic center using its individual determinations
in the last decade by independent methods have been performed by
now. For example, $R_0=8.0±0.2$~kpc (Vall\'ee 2017a),
$R_0=8.4\pm0.4$ kpc (de Grijs and Bono 2017), or $R_0=8.0\pm0.15$
kpc (Camarillo et al. 2018). Based on these reviews, in this paper
we adopted $R_0=8.0\pm0.15$~kpc.

The influence of the spiral density wave in the radial $(V_R)$ and
residual tangential $(\Delta V_{circ})$ velocities is periodic
with an amplitude of $\sim$6--10 km s$^{-1}$. According to the
linear theory of density waves (Lin and Shu 1964), it is described
by the following relations:
 \begin{equation}
 \begin{array}{lll}
       V_R =-f_R \cos \chi,\\
 \Delta V_{circ}= f_\theta \sin\chi,
 \label{DelVRot}
 \end{array}
 \end{equation}
where
 \begin{equation}
 \chi=m[\cot(i)\ln(R/R_0)-\theta]+\chi_\odot
 \end{equation}
is the phase of the spiral density wave ($m$ is the number of
spiral arms, $i$ is the pitch angle of the spiral pattern, and
$\chi_\odot$ is the Sun’s radial phase in the spiral density
wave); $f_R$ and $f_\theta$ are the amplitudes of the radial and
tangential velocity perturbations, which are assumed to be
positive. As an analysis of the present day highly accurate data
showed, the periodicities associated with the spiral density wave
also manifest themselves in the vertical velocities $W$ (Bobylev
and Bajkova 2015; Rastorguev et al. 2017).

We apply a modified spectral analysis (Bajkova and Bobylev 2012)
to study the periodicities in the velocities  $V_R$ and $\Delta
V_{circ}$. The wavelength $\lambda$ (the distance between adjacent
spiral arm segments measured along the radial direction) is
calculated from the relation
\begin{equation}
 \frac{2\pi R_0}{\lambda}=m\cot(i).
 \label{a-04}
\end{equation}
Let there be a series of measured velocities $V_{R_n}$ (these can
be both radial $(V_R)$ and tangential $(\Delta V_{circ})$
velocities), $n=1,\dots,N$, where $N$ is the number of objects.
The objective of our spectral analysis is to extract a periodicity
from the data series in accordance with the adopted model
describing a spiral density wave with parameters $f,$
$\lambda$~(or $i)$ and $\chi_\odot$.

Having taken into account the logarithmic behavior of the spiral
density wave and the position angles of the objects $\theta_n$,
our spectral (periodogram) analysis of the series of velocity
perturbations is reduced to calculating the square of the
amplitude (power spectrum) of the standard Fourier transform
(Bajkova and Bobylev 2012):
\begin{equation}
 \bar{V}_{\lambda_k} = \frac{1} {N}\sum_{n=1}^{N} V^{'}_n(R^{'}_n)
 \exp\biggl(-j\frac {2\pi R^{'}_n}{\lambda_k}\biggr),
 \label{29}
\end{equation}
where $\bar{V}_{\lambda_k}$ is the $k$th harmonic of the Fourier
transform with wavelength $\lambda_k=D/k$, $D$ is the period of
the series being analyzed,
 \begin{equation}
 \begin{array}{lll}
 R^{'}_{n}=R_0\ln(R_n/R_0),\\
 V^{'}_n(R^{'}_n)=V_n(R^{'}_n)\times\exp(jm\theta_n).
 \label{21}
 \end{array}
\end{equation}
The sought-for wavelength $\lambda$ corresponds to the peak value
of the power spectrum $S_{peak}.$ The pitch angle of the spiral
density wave is derived from Eq. (9). We determine the
perturbation amplitude and phase by fitting the harmonic with the
wavelength found to the observational data. The following relation
can also be used to estimate the perturbation amplitude:
 \begin{equation}
 f_R(f_\theta)=2\times\sqrt{S_{peak}}.
 \label{Speak}
 \end{equation}
Thus, our approach consists of two steps: (i) the construction of
a smooth Galactic rotation curve and (ii) a spectral analysis of
the radial $(V_R)$ and residual tangential $(\Delta V_{circ})$
velocities. This method was applied by Bobylev and Bajkova (2012,
2013, 2015, 2018) to study the kinematics of young Galactic
objects.

 \subsubsection*{Monte Carlo Simulations}
We use Monte Carlo simulations to estimate the errors in the
parameters of the spiral density wave being determined. In
accordance with this method, we generate $M$ independent
realizations of data on the parallaxes and velocities of objects
with their random measurement errors that are known to us.

We assume that the measurement errors of the data are distributed
normally with a mean equal to the nominal value and a dispersion
equal to $\sigma_l={error}_l, l=1,\dots,N_d$, where $N_d$ is the
number of data and ${error}_l$ denotes the measurement error of a
single measurement with number $l$ (one sigma). Each element of a
random realization is formed independently by adding the nominal
value of the measured data with number $l$ and the random number
generated according to a normal law with zero mean and dispersion
$\sigma_l.$ Note that the latter is limited from above by
3$\sigma_l.$

Each random realization of data with number $j$ ($j=1,\dots,M$)
generated in this way is then processed according to the algorithm
described above to determine the sought-for parameters $f_R^j,
\lambda^j, \chi_\odot^j$. The mean values of the parameters and
their dispersions are then determined from the derived sequences
of estimates: $m_{f_R}\pm \sigma_{f_R}, m_{\lambda}\pm
\sigma_{\lambda}, m_{\chi_\odot}\pm \sigma_{\chi_\odot}$. The
statistical parameters of the spiral density wave pitch angle $i$
can be determined using Eq. (9): $m_{i}\pm \sigma_{i}$.

 \section*{RESULTS}
The system of conditional equations (3)--(5) is solved by the
least-squares method with weights of the form $w_r=S_0/\sqrt
{S_0^2+\sigma^2_{V_r}},$
 $w_l=S_0/\sqrt {S_0^2+\sigma^2_{V_l}}$ and
 $w_b=S_0/\sqrt {S_0^2+\sigma^2_{V_b}},$ where
$S_0$ is the ``cosmic'' dispersion, $\sigma_{V_r}, \sigma_{V_l},
\sigma_{V_b}$ are the dispersions of the corresponding observed
velocities. $S_0$ is comparable to the root-mean-square residual
$\sigma_0$ (the error per unit weight) in solving the conditional
equations (3)--(5). We adopted $S_0=8$~km s$^{-1}$ when analyzing
the sample of young OSCs and $S_0=11$~km s$^{-1}$ for the sample
of older OSCs. The system of equations (3)--(5) was solved in
several iterations using the 3$\sigma$ criterion to eliminate the
OSCs with large residuals.

{\bf Method I.} The first method consists in seeking a solution
based on such OSCs for which the space velocities $U,V,W$ can be
calculated. First, based on the sample of 211 relatively young
$(\log t<8)$ OSCs, we obtained a solution of the system of
conditional equations (3)--(5) from the original data, i.e.,
without correcting the parallaxes. The following kinematic
parameters were found in this approach:
 \begin{equation}
 \label{solution-00}
 \begin{array}{lll}
 (U_\odot,V_\odot,W_\odot)=(7.63,11.72,8.93)\pm(0.60,0.74,0.61)~\hbox{km s$^{-1}$},\\
      \Omega_0=~28.34\pm0.37~\hbox{km s$^{-1}$ kpc$^{-1}$},\\
  \Omega^{'}_0=-3.832\pm0.090~\hbox{km s$^{-1}$ kpc$^{-2}$},\\
 \Omega^{''}_0=~0.851\pm0.073~\hbox{km s$^{-1}$ kpc$^{-3}$}.
 \end{array}
 \end{equation}
In this solution the error per unit weight is $\sigma_0=8.5$~km
s$^{-1}$.

The next solution of the conditional equations (3)--(5) was
obtained with the corrected OSC parallaxes by applying the
correction $\Delta\pi=0.050$~mas. In this case, the following
kinematic parameters were found:
 \begin{equation}
 \label{solution-1}
 \begin{array}{lll}
 (U_\odot,V_\odot,W_\odot)=(7.36,12.15,8.22)\pm(0.57,0.72,0.57)~\hbox{km s$^{-1}$},\\
      \Omega_0 =~28.79\pm0.39~\hbox{km s$^{-1}$ kpc$^{-1}$},\\
  \Omega^{'}_0 =-3.999\pm0.091~\hbox{km s$^{-1}$ kpc$^{-2}$},\\
 \Omega^{''}_0 =~0.921\pm0.096~\hbox{km s$^{-1}$ kpc$^{-3}$}.
 \end{array}
 \end{equation}
In this solution the error per unit weight is $\sigma_0=7.9$ km
s$^{-1}$. For the adopted $R_0=8.0\pm0.15$ kpc the linear Galactic
rotation velocity $(V_0=|R_0\Omega_0|)$ is $V_0=230\pm6$ km
s$^{-1}$, while the Oort constants
 $(A=-0.5\Omega'_0R_0$ and $B=\Omega_0+A)$ take the
following values:
 $A=16.00\pm0.37$ km s$^{-1}$ kpc$^{-1}$ and
 $B=-12.79\pm0.53$ km s$^{-1}$ kpc$^{-1}$.

{\bf Method II.} In this approach we exploit all potentialities of
the available data. The clusters with the proper motions,
line-of-sight velocities, and distances give all three equations
(3)--(5), while the clusters for which only the proper motions are
available give only two equations, (4) and (5). We solve this
system of equations simultaneously.

We apply this method to analyze OSCs younger than 1~Gyr $(\log
t<9).$ For this purpose, we divided the sample into two parts: 326
relatively young $(\log t<8)$ OSCs and 481 older $(8<\log t<9)$
OSCs.

Based on the sample of young $(\log t<8)$ OSCs, we found the
following kinematic parameters:
 \begin{equation}
 \label{solution-2}
 \begin{array}{lll}
 (U_\odot,V_\odot,W_\odot)=(7.88,11.17,8.28)\pm(0.48,0.63,0.45)~\hbox{km s$^{-1}$},\\
      \Omega_0 =~29.34\pm0.31~\hbox{km s$^{-1}$ kpc$^{-1}$},\\
  \Omega^{'}_0 =-4.012\pm0.074~\hbox{km s$^{-1}$ kpc$^{-2}$},\\
 \Omega^{''}_0 =~0.779\pm0.062~\hbox{km s$^{-1}$ kpc$^{-3}$},
 \end{array}
 \end{equation}
where the error per unit weight is $\sigma_0=7.9$ km s$^{-1}$, the
Galactic rotation velocity is $V_0=235\pm5$ km s$^{-1}$, and the
Oort constants are $A=16.05\pm0.30$ km s$^{-1}$ kpc$^{-1}$ and
$B=-13.29\pm0.43$ km s$^{-1}$ kpc$^{-1}$. Basically, this solution
is an improvement of the solution (14), because 115 more OSCs for
which only the parallaxes and proper motions are available were
added here to the 211 OSCs used in seeking the solution (14).

Based on the sample of 481 older $(8<\log t<9)$ OSCs, we found the
following kinematic parameters:
 \begin{equation}
 \label{solution-3}
 \begin{array}{lll}
 (U_\odot,V_\odot,W_\odot)=(8.58,11.10,7.54)\pm(0.61,0.76,0.52)~\hbox{km s$^{-1}$},\\
      \Omega_0 =~28.42\pm0.39~\hbox{km s$^{-1}$ kpc$^{-1}$},\\
  \Omega^{'}_0 =-3.972\pm0.097~\hbox{km s$^{-1}$ kpc$^{-2}$},\\
 \Omega^{''}_0 =~0.642\pm0.061~\hbox{km s$^{-1}$ kpc$^{-3}$},
 \end{array}
 \end{equation}
where the error per unit weight is $\sigma_0 = 11.1$ km s$^{-1}$,
the Galactic rotation velocity is $V_0=227\pm5$ km s$^{-1}$, and
the Oort constants are
 $A= 15.89\pm0.39$ km s$^{-1}$ kpc$^{-1}$ and
 $B=-12.54\pm0.55$ km s$^{-1}$ kpc$^{-1}$.

\begin{figure}[t]
{\begin{center}
   \includegraphics[width=0.9\textwidth]{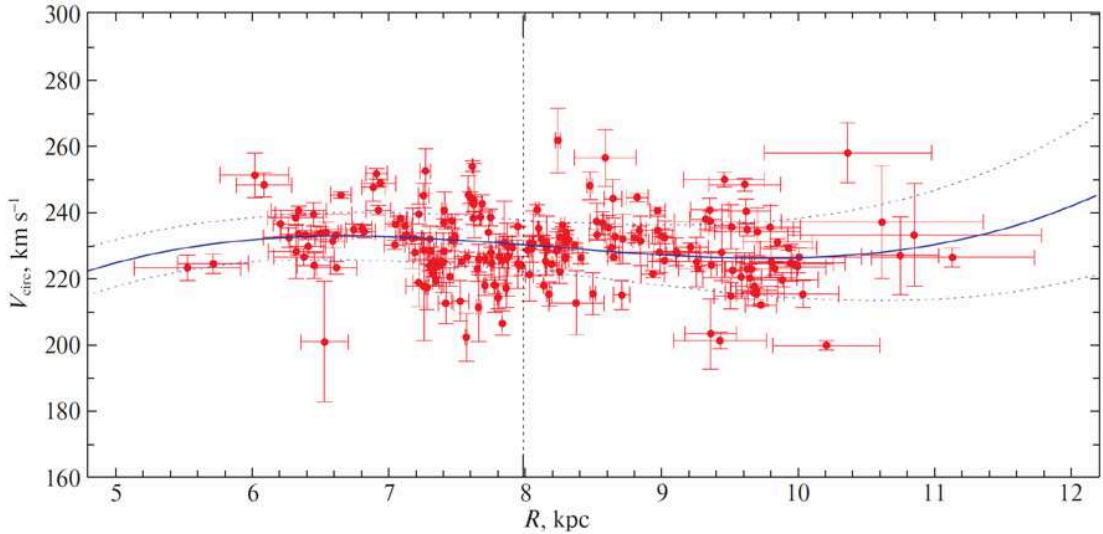}
 \caption{
Circular velocities of young OSCs versus Galactocentric distance.
The Galactic rotation curve constructed according to the solution
(14) with 1$\sigma$ confidence intervals is presented; the
vertical dotted line marks the Sun's position.
  } \label{f-Rotat}
\end{center}}
\end{figure}
\begin{figure}[t]
{\begin{center}
  \includegraphics[width=0.45\textwidth]{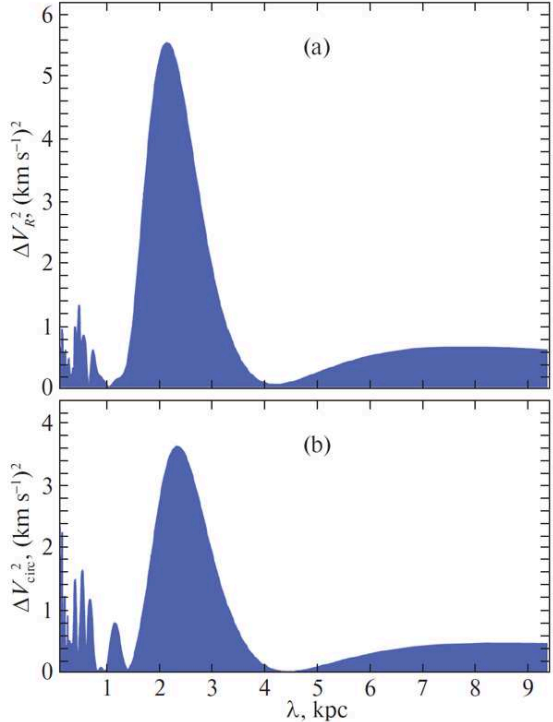}
 \caption{Power spectra of the radial (a) and residual tangential (b) velocities for young OSCs.}
\end{center}}
\end{figure}
\begin{figure}[t]
{\begin{center}
  \includegraphics[width=0.45\textwidth]{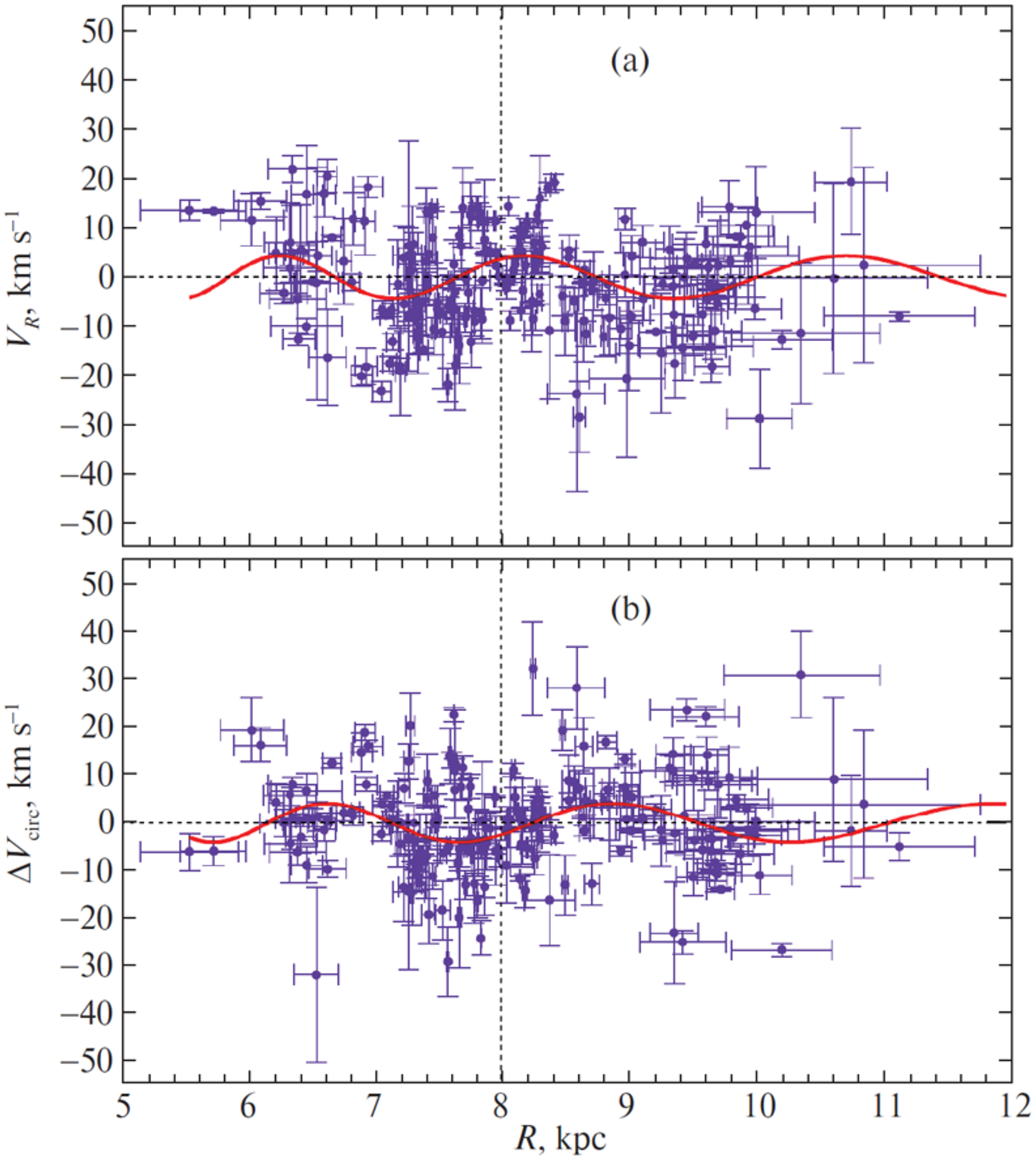}
 \caption{
Radial (a) and residual tangential (b) velocities of young OSCs
versus Galactocentric distance; the vertical dotted line marks the
Sun's position.
  } \label{f-Rest}
\end{center}}
\end{figure}

 \subsection*{Velocity Perturbations from the Density Wave}
In Fig 3 the circular velocities of OSCs are plotted against the
Galactocentric distance; the Galactic rotation curve constructed
according to the solution (15) is presented. As can be seen from
the figure, the residual velocities have a low dispersion; a
periodicity with a length of about 2 kpc is clearly visible.

Based on the deviation from the Galactic rotation curve (15), we
calculated the residual circular velocities $\Delta V_{circ}$.
Based on the series of radial $(V_R)$ and residual tangential
($\Delta V_{circ}$) velocities for this sample of OSCs, we found
the parameters of the Galactic spiral density wave by applying a
periodogram analysis. The amplitudes of the radial and tangential
velocity perturbations are
 $f_R = 4.7\pm1.0$ km s$^{-1}$ and
 $f_\theta= 3.8\pm1.2$ km s$^{-1}$, respectively.

Figure 4 shows the power spectra of the OSC velocities. It is
clearly seen from this figure that the peaks of the distribution
lie almost at the same $\lambda$ in both cases. Indeed, the
perturbation wavelengths are $\lambda_R=2.2\pm0.5$ kpc
($i=-10\pm2^\circ$) and $\lambda_\theta=2.3\pm0.5$ kpc
($i=-11\pm2^\circ$) for the adopted four-armed spiral pattern
$(m=4).$

Figure 5 presents the radial and residual tangential velocities of
OSCs. It is clearly seen that the periodic curves in Figs. 5a and
5b go with a phase shift of $90^\circ$. We measure the Sun's phase
in the spiral density wave $\chi_\odot$ from the
Carina--Sagittarius arm ($R\sim7$ kpc); in our case, its value is
very close to $-120\pm10^\circ.$

A number of OSCs that deviate significantly from the overall
pattern can be seen in Figs. 3 and 5. For example, the cluster
Stock 16 ($R=6.5$ kpc) has a large deviation from the rotation
curve, $\Delta V_{circ}=-32\pm18$ km s$^{-1}$. One more OSC that
does not ``march in step'', NGC 2453 ($R=10.4$ kpc), has $\Delta
V_{circ}=31\pm9$ km s$^{-1}$. Both these clusters have large
relative parallax errors, $\sigma_\pi/\pi=30\%$ for Stock 16 and
$\sigma_\pi/\pi=22\%$ for NGC 2453. Whereas Stock 16 is fairly
young, $\log t=6.78,$ NGC 2453 is older, $\log t=7.86.$ Note that
both these clusters and several more OSCs with smaller random
errors in the velocities $V_R$ and $V_{circ}$ were rejected
according to the 3$\sigma$ criterion when seeking the solutions
(13)--(15).

 \section*{DISCUSSION}
Bobylev et al. (2016) performed a kinematic analysis of OSCs from
the MWSC catalogue (Kharchenko et al. 2013) using photometric
distance estimates. First of all, it should be noted that the
distribution of the sample of young OSCs with trigonometric
parallaxes on the Galactic $XY$  plane (Fig. 2) visually agrees
much better with the model of a spiral pattern than does their
distribution that was derived using photometric distance estimates
(see Fig. 1 in Bobylev et al. (2016)).

The error per unit weight $\sigma_0$ that we find when solving the
conditional equations (3)--(5) characterizes the residual velocity
dispersion for OSCs averaged over three directions. The residual
velocity dispersion for hydrogen clouds in the Galactic disk is
known to be $\sim$5 km s$^{-1}$. The residual velocity dispersion
for OB stars lies in the range 8--10 km s$^{-1}$; the analogous
velocity dispersion for Cepheids is $\sim$14 km s$^{-1}$. One
might expect the velocity dispersion for young OSCs to be close to
that for OB stars. In the solutions (14) and (15) we found
$\sigma_0=7.9$ km s$^{-1}$, which agrees excellently with the
expected value. Therefore, it is surprising that when analyzing
the youngest OSCs from the MWSC catalogue (Kharchenko et al.
2013), $\sigma_0$ is 15.7 km s$^{-1}$ in Bobylev et al. (2016).
This can be explained by the fact that the errors of the stellar
proper motions taken from the PPMXL catalogue (R\"oser et al.
2010), where their values lie in the range 4--10 mas yr$^{-1}$,
i.e., exceed the random errors of the Gaia DR2 stellar proper
motions by two orders of magnitude, are great. The errors of the
photometric distances also make their contribution.

The results of the solution (16) are also of indubitable interest.
The increase in $\sigma_0$ to 11.1 km s$^{-1}$ is related to the
growth of the velocity dispersions with increasing stellar age
(disk heating), with the contribution of the purely measurement
errors being negligible here. For example, for OSCs with a close
age Bobylev et al. (2016) found $\sigma_0=21$ km s$^{-1}$.

Based on a sample of 209 young $(\log t<7.7)$ OSCs from the MWSC
catalogue, Bobylev et al. (2016) found the following solar
velocity components:
$(U_\odot,V_\odot,W_\odot)=(9.7,11.2,6.2)\pm(1.1,1.4,1.1)$ km
s$^{-1}$ and parameters of the Galactic rotation curve:
 $\Omega_0=28.60\pm0.81$~km s$^{-1}$ kpc$^{-1}$,
 $\Omega^{'}_0=-4.04\pm0.16$~km s$^{-1}$ kpc$^{-2}$ and
 $\Omega^{''}_0=0.19\pm0.13$~km s$^{-1}$ kpc$^{-3}$ ($R_0=8.3\pm0.2$ kpc was adopted).
We can see that in the solution (14), at the same number of OSCs,
the errors in the parameters being determined are smaller
approximately by a factor of 2.

Thus, in this paper we used virtually the same line-of-sight
velocities of OSCs as those in Bobylev et al. (2016), but
completely different distances and proper motions of OSCs. As a
result, we obtained reliable ($\sigma_0$ is small), new Galactic
parameters in the solutions (14) and (15).

Having analyzed the proper motions and parallaxes for a local
sample of 304267 main-sequence stars for the Gaia DR1 catalogue,
Bovy (2017) obtained the following Oort parameters:
 $A=15.3\pm0.5$ km s$^{-1}$ kpc$^{-1}$ and $B=-11.9\pm0.4$ km s$^{-1}$ kpc$^{-1}$,
based on which he estimated the angular velocity of Galactic
rotation $\Omega_0=27.1\pm0.5$ km s$^{-1}$ kpc$^{-1}$ and the
corresponding linear velocity $V_0=219\pm4$ km s$^{-1}$.

Based on 130 masers with measured VLBI trigonometric parallaxes,
Rastorguev et al. (2017) found the solar velocity components
$(U_\odot,V_\odot)=(11.40,17.23)\pm(1.33,1.09)$ km s$^{-1}$ and
the following parameters of the Galactic rotation curve:
$\Omega_0=28.93\pm0.53$~km s$^{-1}$ kpc$^{-1},$
 $\Omega^{'}_0=-3.96\pm0.07$~km s$^{-1}$ kpc$^{-2},$
 $\Omega^{''}_0=0.87\pm0.03$~km s$^{-1}$ kpc$^{-3}$ and $V_0=243\pm10$ km s$^{-1}$ (for
 $R_0=8.40\pm0.12$~kpc found).

Based on a sample of 495 OB stars with proper motions from the
Gaia DR2 catalogue, Bobylev and Bajkova (2018) found the following
kinematic parameters:
$(U,V,W)_\odot=(8.16,11.19,8.55)\pm(0.48,0.56,0.48)$ km s$^{-1}$,
 $\Omega_0=28.92\pm0.39$~km s$^{-1}$ kpc$^{-1},$
  $\Omega^{'}_0=-4.087\pm0.083$~km s$^{-1}$ kpc$^{-2},$
 $\Omega^{''}_0=0.703\pm0.067$~km s$^{-1}$ kpc$^{-3}$ and
$V_0=231\pm5$~km s$^{-1}$ (for the adopted $R_0=8.0\pm0.15$ kpc).
We conclude that the kinematic parameters found in the solutions
(14) and (15) are in good agreement with the results of an
analysis of the present-day data obtained by Bovy (2017),
Rastorguev et al. (2017), and Bobylev and Bajkova (2018). Judging
by the level of random errors in the parameters being determined,
the solution (15) is one of the best at present. It is slightly
inferior in parameter $\Omega^{''}_0$ (a large radius of the
neighborhood is required here) only to the solution obtained by
Rastorguev et al. (2017) based on a sample of masers with VLBI
parallaxes.

The parameters of the spiral density wave. The mean pitch angle of
the global four-armed spiral pattern in our Galaxy
$i=-13.6\pm0.4^\circ$ is given in the review by Vall\'ee (1917b).
Then, for $m=4$ and $R_0=8.0$~kpc $\lambda=3.0$ kpc follows from
Eq. (9). We can see that the analysis of our sample of young OSCs
gives a lower value of $\lambda$ and, accordingly, a smaller pitch
angle $|i|:10-11^\circ.$

Having analyzed the spatial distribution of a large sample of
classical Cepheids, Dambis et al. (2015) estimated the pitch angle
of the spiral pattern, $i=-9.5^\circ\pm0.1^\circ$, and the Sun's
phase, $\chi_\odot=-121^\circ\pm3^\circ$, for the four-armed
spiral pattern.

On the other hand, having analyzed maser sources with VLBI
parallaxes, Rastorguev et al. (2017) found
$i=-10.4^\circ\pm0.3^\circ$ and
$\chi_\odot=-125^\circ\pm10^\circ,$ which is in good agreement
with our results. The amplitude of the radial velocity
perturbations $f_R$ is typically 6--10 km s$^{-1}$ from masers
(Rastorguev et al. 2017), OB stars (Bobylev and Bajkova 2015,
2018), or Cepheids (Bobylev and Bajkova 2012). For a more reliable
determination of the spiral density wave parameters, it is
necessary to expand the OSC sample to cover a larger region of the
Galaxy.

 \section*{CONCLUSIONS}
Thus, based on published data, we selected a sample of OSCs with
proper motions and parallaxes from the Gaia DR2 catalogue. The
catalogue by Cantat-Gaudin et al. (2018) served as a basis for
this purpose. The MWSC catalogue (Kharchenko et al. 2013) served
as the main source of line-of-sight velocities; for several OSCs
the line-of-sight velocities were taken from the Gaia DR2
catalogue. This sample includes a total of 925 OSCs of various
ages with relative parallax errors less than 30\%.

The sample of 326 youngest OSCs with an age $\log t<8$ was studied
in detail. All these clusters are no farther than 5 kpc away from
the Sun and no higher than 300 pc above the Galactic plane. They
were used to redetermine the Galactic rotation parameters and the
parameters of the spiral density wave.

Following the latest results of an analysis of the zero point for
the Gaia DR2 distance scale, we calculated the distances to OSCs
by adding the correction $\Delta\pi=0.050$ mas to the original
mean values of their parallaxes.

As a result, we found the following parameters of the angular
velocity of Galactic rotation:
 $\Omega_0 =29.34\pm0.31$~km s$^{-1}$ kpc$^{-1},$
 $\Omega^{'}_0=-4.012\pm0.074$~km s$^{-1}$ kpc$^{-2}$ and
 $\Omega^{''}_0=0.779\pm0.062$~km s$^{-1}$ kpc$^{-3}$;
here the circular rotation velocity of the solar neighborhood
around the Galactic center is $V_0=235\pm5$ km s$^{-1}$ for the
adopted distance $R_0=8.0\pm0.15$ kpc.

The influence of the Galactic spiral density wave was detected
both in the spatial distribution and in the velocities of the
sample under study. A spectral analysis of the radial and residual
tangential velocities for young OSCs showed excellent agreement in
the perturbation wavelengths found independently for each type of
velocities, $\lambda_R=2.2\pm0.5$~kpc and
$\lambda_\theta=2.3\pm0.5$ kpc. For the four-armed spiral pattern
($m=4$ and the adopted $R_0$) a pitch angle $i\sim-10^\circ$
corresponds to these values. The Sun's phase in the spiral density
wave is close to $\chi_\odot=-120^\circ\pm10^\circ$. The
amplitudes of the radial and tangential velocity perturbations are
$f_R=4.7\pm1.0$~km s$^{-1}$ and $f_\theta=3.8\pm1.2$ km s$^{-1}$,
respectively.

We also considered a sample of 481 older $(\log t:8-9)$ OSCs.
These OSCs were shown to rotate more slowly, with a velocity
$V_0=227\pm5$ km s$^{-1}$. The parameters of the spiral density
wave were not determined for this sample.

 \subsubsection*{ACKNOWLEDGMENTS}
We are grateful to the referees for their useful remarks that
contributed to an improvement of the paper. This work was
supported in part by Basic Research Program P--28 of the Presidium
of the Russian Academy of Sciences, the subprogram ``Cosmos:
Studies of Fundamental Processes and their Interrelations''.

 \bigskip
 \bigskip\medskip{\bf REFERENCES}
{\small

1. F. Arenou, X. Luri, C. Babusiaux, C. Fabricius, A. Helmi, T.
Muraveva, A. C. Robin, F. Spoto, et al. (Gaia Collab.), Astron.
Astrophys. 616, 17 (2018).

2. C. Babusiaux, F. van Leeuwen, M. A. Barstow, C. Jordi, A.
Vallenari, A. Bossini, A. Bressan, T. Cantat-Gaudin, et al. (Gaia
Collab.), Astron. Astrophys. 616, 10 (2018).

3. A. T. Bajkova and V. V. Bobylev, Astron. Lett. 38, 549 (2012).

4. G. Beccari, H. M. J. Boffin, T. Jerabkova, N. J. Wright, V. M.
Kalari, G. Carraro, G. De Marchi, and W.-J. de Wit, Mon. Not. R.
Astron. Soc. 481, L11 (2018).

5. V. V. Bobylev and A. T. Bajkova, Astron. Lett. 38, 638 (2012).

6. V. V. Bobylev and A. T. Bajkova, Astron. Lett. 39, 532 (2013).

7. V. V. Bobylev and A. T. Bajkova, Mon. Not. R. Astron. Soc. 437,
1549 (2014).

8. V. V. Bobylev and A. T. Bajkova, Astron. Lett. 41, 473 (2015).

9. V. V. Bobylev and A. T. Bajkova, Astron. Lett. 42, 1 (2016).

10. V. V. Bobylev, A. T. Bajkova, and K. S. Shirokova, Astron.
Lett. 42, 721 (2016).

11. V. V. Bobylev and A. T. Bajkova, Astron. Lett. 44, 675 (2018).

12. V. V. Bobylev, Astron. Lett. 45, 10 (2019).

13. J. Bovy, Mon. Not. R. Astron. Soc. 468, L63 (2017).

14. A. G. A. Brown, A. Vallenari, T. Prusti, de Bruijne, C.
Babusiaux, C. A. L. Bailer-Jones, M. Biermann, D.W. Evans, et al.
(Gaia Collab.), Astron. Astrophys. 616, 1 (2018).

15. T. Camarillo, M. Varun, M. Tyler, and R. Bharat, Publ. Astron.
Soc. Pacif. 130, 4101 (2018).

16. T. Cantat-Gaudin, C. Jordi, A. Vallenari, A. Bragaglia, L.
Balaguer-Nu\'nez, C. Soubiran, et al., Astron. Astrophys. {\bf
618}, 93 (2018).

17. L. Casamiquela, R. Carrera, C. Jordi, L. Balaguer-Nu\'nez, E.
Pancino, S. L. Hidalgo, C. E. Martinez-Vazquez, S. Murabito, et
al., Mon. Not. R. Astron. Soc. 458, 3150 (2016).

18. C. Conrad, R.-D. Scholz, N. V. Kharchenko, A. E. Piskunov, E.
Schilbach, S. R\"oser, C. Boeche, G. Kordopatis, et al., Astron.
Astrophys. 562, 54 (2014).

19. A. K. Dambis, L. N. Berdnikov, Yu. N. Efremov, A. Yu. Kniazev,
A. S. Rastorguev, E. V. Glushkova, V. V. Kravtsov, D. G. Turner,
D. J. Majaess, and R. Sefako, Astron. Lett. 41, 489 (2015).

20. W. S. Dias, H. Monteiro, J. R. D. L\'epine, R. Prates, C. D.
Gneiding, and M. Sacchi, Mon. Not. R. Astron. Soc. 481, 3887
(2018).

21. E. Franciosini, G. G. Sacco, R. D. Jeffries, F. Damiani, V.
Roccatagliata, D. Fedele, and S. Randich, Astron. Astrophys. 616,
12 (2018).

22. R. de Grijs and G. Bono, Astrophys. J. Suppl. Ser. 232, 22
(2017).

23. N. V. Kharchenko, A. E. Piskunov, E. Schilbach, S. R\"oser,
and R.-D. Scholz, Astron. Astrophys. 558, 53 (2013).

24. M. Kounkel, K. Covey, G. Su\'arez, C. Roman-Zu\'niga, J.
Hernandez, K. Stassun, K. O. Jaehnig, E. D. Feigelson, et al.,
Astron. J. 156, 84 (2018).

25. M. A. Kuhn, L. A. Hillenbrand, A. Sills, E. D. Feigelson, and
K. V. Getman, Astrophys. J. 870, 32 (2019).

26. C. C. Lin and F. H. Shu, Astrophys. J. 140, 646 (1964).

27. L. Lindegren, J. Hernandez, A. Bombrun, S. Klioner, U.
Bastian, M. Ramos-Lerate, A. de Torres, H. Steidelmuller, et al.
(Gaia Collab.), Astron. Astrophys. 616, 2 (2018).

28. J. C. Mermilliod, M. Mayor, and S. Udry, Astron. Astrophys.
485, 303 (2008).

29. M. H. Pinsonneault, Y. P. Elsworth, J. Tayar, A. Serenelli, D.
Stello, J. Zinn, S. Mathur, R. Garcia, et al., Astrophys. J.
Suppl. Ser. {\bf 239}, 32 (2018).

30. A. S. Rastorguev, M. V. Zabolotskikh, A. K. Dambis, N. D.
Utkin, V. V. Bobylev, and A. T. Bajkova, Astrophys. Bull. 72, 122
(2017).

31. A. G. Riess, S. Casertano, W. Yuan, L. Macri, B. Bucciarelli,
M. G. Lattanzi, J. W. MacKenty, J. B. Bowers, et al., Astrophys.
J. 861, 126 (2018).

32. V. Roccatagliata, G. G. Sacco, E. Franciosini, and S. Randich,
Astron. Astrophys. 617, L4 (2018).

 33. S. R\"oser, M. Demleitner, and E. Schilbach, Astron. J. 139, 2440 (2010).

34. C. Soubiran, T. Cantat-Gaudin, M. Romero-Gomez, L.
Casamiquela, C. Jordi, A. Vallenari, T. Antoja, L.
Balaguer-Nu\'nez, et al., Astron. Astrophys. {\bf 619}, 155
(2018).

35. K. G. Stassun and G. Torres, Astrophys. J. 862, 61 (2018).

36. J. P. Vall\'ee, Astrophys. Space Sci. 362, 79 (2017a).

37. J. P. Vall\'ee, New Astron. Rev. 79, 49 (2017b).

38. Y. Xu, S. B. Bian, M. J. Reid, J. J. Li, B. Zhang, Q. Z. Yan,
T. M. Dame, K. M. Menten, et al., Astron. Astrophys. 616, L15
(2018).

39. L. N. Yalyalieva, A. A. Chemel’, E. V. Glushkova, A. K.
Dambis, and A. D. Klinichev, Astrophys. Bull. 73, 335 (2018).

 40. E. Zari, H. Hashemi, A. G. A. Brown, K. Jardine, and P. T. de Zeeuw,
   Astron. Astrophys. {\bf 620}, 172 (2018).

 41. J. C. Zinn, M. H. Pinsonneault, D. Huber, and D. Stello,
  arXiv: 1805.02650 (2018).
  }
  \end{document}